\documentclass[11pt]{article}
\usepackage{pst-all}
\usepackage{pstricks}
\usepackage{pstcol,pst-fill,pst-grad}
\usepackage{amsfonts}
\usepackage{graphicx}
\usepackage{amsmath}

\newtheorem{theorem}{Theorem}
\newtheorem{acknowledgement}[theorem]{Acknowledgement}

\makeatletter \@addtoreset{equation}{section}

\newcommand{\be}{\begin{equation}}
\newcommand{\ee}{\end{equation}}
\newcommand{\bea}{\begin{eqnarray}}
\newcommand{\eea}{\end{eqnarray}}
\input{tcilatex}
\begin{document}

\date{}
\title{ \textbf{\ 2D Material Structures and Discrete Symmetries of Periodic
Polygons\thanks{
This work is dedicated to the memory of Professor Ahmed Intissar
(1951-2017). One of the authors (AB) has learned from him many things about
symmetry. } }\\
\textbf{\ } }
\author{ Adil Belhaj$^1$, Salah Eddine Ennadifi$^{2}$ \hspace*{-8pt} \\
{\small $^{1}$ Equipe des sciences de la mati\`{e}re et du Rayonnement,
ESMAR, Facult\'{e} des sciences}\\
{\small Universit\'{e} Mohammed V, Rabat, Morocco} \\
{\small $^2$ Facult\'{e} des sciences de Rabat, Universit\'{e} Mohammed V,
Rabat, Morocco}}
\maketitle

\begin{abstract}
In this work, we reconsider the study of 2D materials involving double
lattice structures associated with periodic polygons. In tessellated
periodic representation, it appears two periodic polygons of $k$ sides of
unequal side lengths at certain angles fixed by the underlying discrete
symmetries. In this way, 2D materials could be engineered by using two
superstructures on the same atomic sheet generated by two length parameters $%
a_{1}$ and $a_{2}$ and rotated by the angle $\phi _{n}^{k}=\frac{n\pi }{k}$,
where $n$ is an arbitrary natural number. These geometrical configurations
could be exploited to engineer 2D materials by doubling the number of the
ordinary unit cell atoms. To support the present conjecture, we establish a
link with Lie symmetries including finite and indefinite ones providing a
room interpretation for $n$.
\end{abstract}


\thispagestyle{empty}

\newpage{}\setcounter{page}{1} \newpage{}

\section{Introduction}

Two-dimensional (2D) materials, composed of semi-conductor atoms, have been
considered as a subject of great interest in condensed matter physics and
related topics. Concretely, many efforts have been devoted to investigate
the electronic structure of such materials using different engineering
methods and various numerical code approaches \cite{1,2,3}. The most treated
ones are graphene, silicene and germanene. It is remarked that graphene,
which is a monolayer of carbon atoms forming 2D hexagonal lattice, has been
the focus of extensive theoretical and experimental investigations, due to
its important physical properties including high carrier mobility and high
electrical and thermal conductivity \cite{4}. For this reason, many 2D
materials have been synthesized and modeled using physical methods and
geometrical approaches associated with non trivial lattice configurations 
\cite{5,6}.

The aim of this work is to contribute to this program by proposing a new
take on the study of 2D materials based on double lattice structures
associated with periodic polygons in the flat geometry. In tessellated
periodic representation, it arises two periodic polygons of $k$ sides, which
will be refereed to as $k$-polygons, of unequal side lengths rotated at
certain angles fixed by the underlying discrete symmetries. This produces
two superstructures on the same atomic sheet generated by two length
parameters $a_{1}$ and $a_{2}$ and rotated by the angle $\phi _{n}^{k}=\frac{%
n\pi }{k}$, where $n$ is an arbitrary natural number. To support the present
proposition, we establish a link with Lie symmetries including finite and
indefinite ones providing a room for $n$. These geometrical configurations
could be exploited to engineer materials by doubling the number of the usual
unit cell atoms. \newline
The paper is organized as follows. In section 2, we give a concise review on
lower dimensional lattice configurations. Section 3 concerns a new take on
2D materials built from periodic $k$-polygons. A connection with Lie
symmetries is discussed in section 4, providing a possible interpretation
for $n$. Section 5 is devoted to concluding remarks and some speculations.

\section{Crystal structures}

\subsection{Lattice symmetry}

Many solid materials involve a crystalline structure that signifies spatial
repetition or periodicity, or more appropriately translation symmetry \cite%
{7}. This is equivalent to say that at every point in the crystal structure
the space behaves the same. In arbitrary dimensions $d$, a lattice is a
regular arrangement of points $M$ at a spatial position $r$ with a
translation $\widehat{T}\left( a_{i}\right) $. This is usually regarded as
an operator which shifts these points $M$ by a displacement amount $a_{i}$
in certain directions. Precisely, any point $M$ located at a lattice
position $r$ can be associated with a vector given by 
\begin{equation}
M\left( r\right) =\sum\limits_{i=1}^{d}m_{i}a_{i}  \label{1}
\end{equation}%
where $m_{i}$ are integer coefficients and $a_{i}$ indicate the primitive
translation vectors or primitive lattice (fundamental) vectors. In this way,
there exists a translation operator $\widehat{T}\left( a_{i}\right) $ 
\begin{equation}
\widehat{T}\left( a_{i}\right) =\sum\limits_{i=1}^{d}t_{i}a_{i}  \label{2}
\end{equation}
shifting the point $M$ by the amount $a_{i}$ such as 
\begin{equation}
\widehat{T}\left( a_{i}\right) M\left( r\right) =\sum\limits_{i=1}^{d}\left(
m_{i}+t_{i}\right) a_{i}=M\left( r+a_{i}\right) ,  \label{3}
\end{equation}%
where $t_{i}$ are arbitrary integers. This operation symmetry generates a
lattice point $M$ at the position $r+a_{i}$. In equations (\ref{1}), (\ref{2}%
) and (\ref{3}) $m_{i}$ and $a_{i}$ are typical $d$-dimensional crystal
structure elements satisfying specific properties in concrete lattices. This
means that all lattice points can be restored by repetition of the basis
vectors $a_{i}$ using the associated symmetry. In connection with solid
state physics, lower dimensional structures can play crucial roles in
building interesting materials.

\subsection{3D Lattices}

Having presented the crystal geometry in terms of the lattice structure \cite%
{8}, we consider now the concrete case of Bravais lattices. In $d=3$, we
assume that there are 3 non-coplanar vectors $a_{i}, \; i=1,2,3$ that
maintain all the properties of the crystal after the shift as a whole by any
of these vectors. As a consequence, any point $N$ in Bravais lattice could
be obtained from  $M$ using translation shifts 
\begin{equation}
N(r)=\sum\limits_{i=1}^{3}\left( m_{i}+t_{i}\right) a_{i}=M\left(
r+a_{i}\right) =\widehat{T}\left( a_{i}\right) M\left( r\right) .  \label{4}
\end{equation}%
In this way, the crystal structure could be understood in terms of Bravais
lattice and the properties of the basis $a_{i}$. It is worth mentioning that
the way of choosing the lattice vectors $a_{i}$ is not unique. However, they
are chosen such that to be primitive. This means that they are chosen in the
lattice as the shortest period, with $a_{2}$ not parallel to $a_{1}$ and $%
a_{3}$ not coplanar to $a_{1}$ and $a_{2}$. The volume delimited by such
primitive vectors defines the primitive unit cell, given by 
\begin{equation}
V_{unit}=(a_{1}.\left\vert a_{2}\wedge a_{3}\right\vert ).  \label{5}
\end{equation}%
In this regard, the lattice vectors $a_{i}$ and the angles between them $%
\theta _{i}$ generate a 6-dimensional parameter space (or moduli space).
This space determines the lattice structure and its symmetries. In 3D, it
turns out that this space could have the following topology 
\begin{equation}
(\mathbb{R}_{+}^{\star })^{3}\times (S^{1})^{3}  \label{05}
\end{equation}%
where $S^{1}$ is the 1-dimensional circle. This indicates that any lattice
structure depends on

\begin{itemize}
\item three length parameters, $a_i$

\item three angular parameters, $\theta_i$.
\end{itemize}

Concretely, the types of lattices can be associated with some particular
regions in such a moduli space. In particular, they differ by the relations
between the lengths $a_{i}$ and the angles $\theta _{i}$. Indeed, there are
14 types of lattices with 7 crystal classes giving rise to several primitive
cells \cite{9}. These 14 lattices correspond to the following lattice
parameter constraints associated with particular regions in the above moduli
space 
\begin{eqnarray}
a_{1} &=&a_{2}=a_{3}\text{ \ with\ \ }\theta _{1}=\theta _{2}=\theta _{3}=90%
{{}^\circ}%
\text{ or}\neq 90%
{{}^\circ}
\notag \\
a_{1} &=&a_{2}\neq a_{3}\text{ \ with\ \ }\theta _{1}=\theta _{2}=\theta
_{3}=90%
{{}^\circ}%
\text{ or }\theta _{1}=\theta _{2}=90%
{{}^\circ}%
\text{and }\theta _{3}=120%
{{}^\circ}
\label{6} \\
a_{1} &\neq &a_{2}\neq a_{3}\text{ \ with\ \ }\theta _{1}=\theta _{2}=\theta
_{3}=90%
{{}^\circ}%
\text{ or}\neq 90%
{{}^\circ}%
\text{, }\theta _{1}=\theta _{3}=90%
{{}^\circ}%
\text{and }\theta _{2}\neq 120%
{{}^\circ}%
.  \notag
\end{eqnarray}%
In 2D, few crystal classes and primitive cells are discussed. The
corresponding 2D crystals are known to have use in many physical
applications, including graphene like-materials.

\subsection{2D Lattices}

In 2D, the above mentioned 14 Bravais lattices in 3D are greatly reduced to
5 ones. This is due to the fact that there should be just two non-colinear
basis vectors $a_{1}$ and $a_{2}$. These two fundamental vectors are needed
to span 2D with the following delimited area 
\begin{equation}
A_{unit}=\left\vert a_{1}\wedge a_{2}\right\vert   \label{7}
\end{equation}%
defining the primitive unit cell. In fact, the total number of possible 2D
Bravais lattices is abased due to the existing translational symmetry given
by integer multiples of the basis vectors. These 5 lattices correspond to
the following parameter constraints 
\begin{eqnarray}
a_{1} &=&a_{2}\text{ \ with\ \ }\theta =90%
{{}^\circ}%
\text{ or}=120%
{{}^\circ}
\notag \\
a_{1} &\neq &a_{2}\text{ \ with\ \ }\theta =90%
{{}^\circ}%
\text{ or}\neq 90%
{{}^\circ}%
.  \label{8}
\end{eqnarray}%
An examination of such structures shows that the symmetry of periodic $k$
-polygons, in the flat plane, has not been exploited in building crystals.
We will see in the following sections that we can go beyond the above
structures by introducing such a powerful tool explored in modern physics
including condensed matter physics. This may offer a new way to engineer non
trivial structures associated with 2D materials.

\section{New take on 2D materials}

In this section, we would like to go beyond the above 2D structures by
reducing the associated moduli space. As in 3D, it is noted that such a
space has the following topology 
\begin{equation}
(\mathbb{R}_{+}^{\star })^{2}\times S^{1}  \label{05}
\end{equation}%
coordinated by

\begin{itemize}
\item two length parameters $a_i$,

\item one angular parameter $\theta$.
\end{itemize}

The last one is associated with the SO(2) Lie group defining rotations in
the plane. At this level, a deep inspection reveals, however, that the
ordinary structures do not carry any explicit symmetry fashion of periodic $k
$-polygon geometries. Here, we try to give a possible extended view dealing
with 2D materials using a geometric approach based on discrete symmetries
derived from the behaviors of periodic $k$-polygons. To be precise, we
explore techniques of $k$-polygon geometries to support the existence of new
structures which could be exploited in the 2D materials physics. Our
analysis will be based on a geometric reconsideration of the above
classification. In fact, the possible structures will correspond to allowed
periodic $k$-polygon geometries recovering the known ones. Before
proceeding, let us first start by giving some arguments. First, the symmetry
of periodic $k$-polygons should be involved in the classification. In this
scenario, the discrete symmetries $Z_{k}$ will be primordial in the
determination of the nature of such 2D structures. Second, the dimension of
the lattice could play a crucial role in the related construction. Third,
the reduced moduli space could be exploited to make links between these
arguments. At first sight, it seems that this link does not work. However,
group theory and Lie symmetries push us to think about a new take on the
above structures. An investigation reveals that the discrete symmetries $%
Z_{k}$ can be exploited to produce new geometrical materials involving
double structures. To show how this works in practice, we could solve the
constraint $a_{2}\neq a_{1}$ by a possible reduction of the associated
moduli space. A priori, there are many ways to handle this constraint. Here,
we propose a possible way giving rise new geometrical structures in 2D.
Combining the above arguments, we can anticipate the following relation 
\begin{equation}
a_{2}=f(k)a_{1}  \label{051}
\end{equation}%
where $f(k)$ is a real function carrying data on the dimension of the
material and the discrete symmetries $Z_{k}$ of periodic $k$-polygons. A
close inspection shows that we can propose the following formulae 
\begin{equation}
a_{2}=2^{n}|\cos (\frac{n\pi }{k})|a_{1}.  \label{fundamentalequation}
\end{equation}%
In this way, the function $f(k)$ takes the form 
\begin{equation}
f(k)=2^{n}|\cos (\frac{n\pi }{k})|,  \label{052}
\end{equation}%
where 2 is now interpreted as the lattice dimension and $n$ is an arbitrary
natural number constrained by 
\begin{equation}
n\neq (p+\frac{1}{2})k,  \label{052}
\end{equation}%
where $p$ is a natural number used to insure non vanishing functions.
However, the factor $\frac{n\pi }{k}$, which generates a cyclic group
representation of ${Z}_{k}$ discrete symmetry, could play an important role
in solid state physics with double $k$-polygonal structures. It turns out
that we can build up two families:

\begin{enumerate}
\item single structure associated with $n=0$ generated by one length
parameter $a=a_1=a_2$ recovering the ordinary structures, appearing in
graphene like-models.

\item double structure corresponding to $n\neq 0$ provided by two different
length parameters $a_1$ and $a_2$.
\end{enumerate}

At this level, one may comment the last family being not obvious. Indeed,
this involves two periodic $k$-polygon geometries producing 2D materials
having the property of being close to ordinary ones associated with the
first family. In this view, each length parameter produces a single
structure with a specific size. The small one is generated by $a_1$ while
the big one is generated by $a_2$ and rotated by the angle $\phi_n^k=\frac{%
n\pi}{k}$. This family can be exploited to engineer materials with a double
structure obtained by doubling the number of the ordinary unit cell atoms.

To provide a complete analysis, it should be interesting to interpret the
natural number $n$. It turns our that this number can find a place in Lie
symmetries explored in many branches of physics including condensed matter
physics.

\section{Connections with rank two Lie symmetries}

In this section, we would like to make contact with Lie symmetries. This may
support the above proposed relation. Indeed, it is recalled that these
symmetries appearing in any physical area including high energy and
condensed matter physics \cite{10,11,12,13}. To see how this link works in
practice, let us first give a concise review on such symmetries. Roughly
speaking, a Lie symmetry $L$ is a vector space together with an
antisymmetric bilinear bracket $[,]$ : $L\times L\rightarrow L$ verifying
certain conditions including the famous Jacobi identity ($%
([x,[y,z]]+[z,[x,y]]+[y,[z,x]]=0)$). Finite Lie symmetries involve a
particular subalgebra called Cartan subalgebra noted $H$ considered as the
maximal toric abelian Lie sub-algebra playing a remarkable r\^{o}le in the
theory of Lie symmetries. Concretely, it is remarked that $L$ could be
decomposed as the direct sum of $H$ and a vector space $E_{\Delta }$ 
\begin{equation}
L=H\oplus E_{\Delta }.
\end{equation}%
In this way, $E_{\Delta }$ reads as 
\begin{equation}
E_{\Delta }=\oplus _{\alpha }L_{\alpha }.
\end{equation}%
It is noted that the space $L_{\alpha }$ can be defined by 
\begin{equation}
L_{\alpha }=\{x\in L|[h,x]=\alpha (x)x\}
\end{equation}%
for $x$ inside $L$ and $\alpha $ ranges over all elements of the dual of $H$%
. In Lie symmetries, $\alpha $ are called roots which will be interpreted as
atom positions in 2D materials. It is well known that a root system $\Delta $
of a Lie symmetry is defined as a subset of an euclidean space $E$
satisfying certain constraints. Indeed, $\Delta $ is a finite set generating 
$E$. It is noted that 0 is not an element of it. $k\alpha $ is a root only
for $k=\pm \alpha $. For any root $\alpha $, $\Delta $ is invariant under
reflection $\sigma _{\alpha }$, where $\sigma _{\alpha }(\beta )=\beta
-<\beta ,\alpha >\alpha $. The last condition concerns the quantity $<\beta
,\alpha >$ which should be an integer.

The root system $\Delta $ carries various information about the
corresponding Lie symmetries. Such information will be crucial in the
present work to handle the number $n$. The link that we are after pushes us
to consider a special class of Lie symmetries involving two simple roots $%
\alpha_1$ and $\alpha_2$. The associated finite Lie symmetries involve a
nice formulae which reads as 
\begin{equation}
\dim L=2+|\Delta |.
\end{equation}
Here, $|\Delta |$ denotes the number of the roots associated with $L$ which
will be considered as the unit cell atom number. However, 2 is the
associated rank which can be interpreted as the dimension of material
lattice on which atoms will be placed. An alternative way to approach these
symmetries is to exploit the Cartan matrices derived from the non trivial
scalar product between two simple roots $\alpha _{1}$ and $\alpha _{2}$. For
such Lie symmetries, the Cartan matrices $A=(A_{ij})$ read as 
\begin{equation}
A=\left( 
\begin{array}{cc}
2 & \langle \alpha _{1},\alpha _{2}\rangle \\ 
\langle \alpha _{2},\alpha _{1}\rangle & 2%
\end{array}%
\right).
\end{equation}
where $\langle \alpha _{1},\alpha _{2}\rangle $ is given in terms of the
euclidean scalar product between $\alpha _{i}$ 
\begin{equation}
\langle \alpha _{i},\alpha _{j}\rangle =2\frac{(\alpha _{i}.\alpha _{j})}{%
(\alpha _{j}.\alpha _{j})}, \qquad i,j=1,2.
\end{equation}%
Moreover, its has been shown that this matrix can be illustrated in a nice
graph called Dynkin graphs, or diagrams, containing two nodes. Like in graph
theory, the diagonal elements correspond to nodes and the non-diagonal
elements are associated with the number of the edges between them. In fact,
the number of the edges between the node $1$ and the node $2$ is given by $%
\langle \alpha _{1},\alpha _{2}\rangle \langle \alpha _{2},\alpha
_{1}\rangle $. In what follows, we combine these Lie structures and the
symmetry of periodic $k$-polygons to present a possible interpretation of $n$%
. Indeed, this interpretation can be derived from a correspondence between
the two simple roots and the geometry of 2D materials derived from periodic $%
k$-polygons, offering a new approach to elaborate the corresponding
materials. The interpretation, that we are after, can be supported by an
interplay between the material unit cell and the root system $\Delta$ of
rank two Lie symmetries. In this way, each atom placed on the periodic $k$%
-polygon unit cell corresponds to a non-zero root of the Lie symmetry in
question. The lattice parameters $a_{i}$ are associated with the length of
the roots 
\begin{equation}
a_{i}\leftrightarrow |\alpha _{i}|,\qquad i=1,2.
\end{equation}%
In this way, the lattice material configuration can be built by using the
fact that the periodic $k$-polygons tessellate the full plane generating
supercell structure materials. This means that it covers a flat geometry by
repeating a single shape, without gaps or overlapping. It turns out that one
may distinguish between two possible geometric configurations depending on
the values of $k$. The first model involves rectangular geometries
associated with $k=4$. However, the second model is described by $k=6$ based
on the hexagonal structures. \newline
To make contact with the above discussion, it is remarked that one should
consider two separate cases $n=0$ and $n\neq 0$. We will see that these
cases are associated with single and double $k$-polygon structures,
respectively.

\subsection{4-polygon structures}

Here, we deal with the periodic 4-polygon case related to the periodic
rectangular geometry. A close examination shows that $n$ could have a room
in the root system. A possible link is to connect $n$ with the scalar
product between simple two simple roots. For 4-polygon structures, a close
examination shows that one should consider the following identification 
\begin{equation}
\langle \alpha _{1},\alpha _{2}\rangle \langle \alpha _{2},\alpha
_{1}\rangle =2n.
\end{equation}%
In order to check this link, let us first consider $n=0$ producing materials
with single rectangular structure associated with 
\begin{equation}
a_{2}=a_{1}\equiv a, \qquad \phi_0^4=0.
\end{equation}%
In fact, the positions of the 4 atoms placed on the 4-polygon unit cell are
associated with four roots of the Lie symmetry constrained by 
\begin{equation}
\langle \alpha _{1},\alpha _{2}\rangle =\langle \alpha _{2},\alpha
_{1}\rangle =0, \qquad |\Delta|=4.
\end{equation}
In this case, the contact with 2D materials is assured by 
\begin{equation}
a=|\alpha _{1}|=|\alpha _{2}|.
\end{equation}%
Taking now $n=1$, we could get materials with a double 4-polygon structure.
This involves two 4-polygons generated by two length parameters $a_{1}$ and $%
a_{2}$ required by 
\begin{equation}
a_{2}=\sqrt{2}a_{1},
\end{equation}%
and rotated by the angle 
\begin{equation}
\phi_1^4=\frac{\pi}{4}.
\end{equation}
The corresponding Lie symmetry is constrained by the following conditions 
\begin{equation}
\langle \alpha _{1},\alpha _{2}\rangle =-2, \quad \langle \alpha _{2},\alpha
_{1}\rangle =-1, \qquad |\Delta|=8, \qquad |\alpha _{2}|=\sqrt{2}|\alpha
_{1}|.  \label{8atoms}
\end{equation}
The associated material structure with the flat geometry can be derived from
the fact 4-polygons tessellate the full plane producing a new crystal by
mixing the $a(1\times 1)$ and $a(\sqrt{2}\times \sqrt{2})$ configurations
rotated by $\phi_1^4=\frac{\pi}{4}.$ In this way, the positions of 8 atoms
are associated with the root system controlled by (\ref{8atoms}).  Inspired
by indefinite Kac-Moody theory, we can do something similar for $n>1$ by
going beyond finite Lie symmetries. It is recalled that the Cartan matrices $%
A$ of extended Lie symmetries should obey the property 
\begin{equation}
det \; A < 0.
\end{equation}
Combing certain results of Lie symmetries, the case $n>1$ could be linked to
the following Cartan matrix  
\begin{equation}
A=%
\begin{pmatrix}
2 & -2n \\ 
-1 & 2%
\end{pmatrix}%
.
\end{equation}
In order to get a possible Dynkin geometry, we need to require 
\begin{equation}
n>2.
\end{equation}
Like in the finite Lie symmetries, we associate to this matrix an extended
Dynkin diagram involving two nodes connected by $2n$ lines. This geometry
produces a double periodic 4-polygon structure involving 8 atoms placed in
the associated unit cell. The configuration contains two periodic 4-polygons
with two different sizes. The small one is generated by $a_1$ while the
second one is engineered by $a_2$ rotated by $\frac{n\pi}{4}$.

\subsection{Links with 6-polygons}

In similar manner, we reconsider the 6-polygon case corresponding to
periodic hexagonal geometry. As in the previous case, $n$ will be linked to
the scalar product of simple roots. Inspired by finite Lie symmetry, we are
now concerned with the following identification 
\begin{equation}
\langle \alpha _{1},\alpha _{2}\rangle \langle \alpha _{2},\alpha
_{1}\rangle =2n+1
\end{equation}
providing odd numbers of edges of the associated Dynkin diagrams.
Considering $n=0$, one can recover 
\begin{equation}
\langle \alpha _{1},\alpha _{2}\rangle \langle \alpha _{2},\alpha
_{1}\rangle =1
\end{equation}%
This is associated with a single hexagonal material having a single
structure having $a_{2}=a_{1}$. Indeed, the positions of the 6 atoms
localized on the 6-polygon unit cell correspond to six roots of the Lie
symmetry constrained by 
\begin{equation}
\langle\alpha _{1},\alpha _{2}\rangle =\langle\alpha _{2},\alpha _{1}\rangle
=-1, \qquad |\Delta|=6.
\end{equation}
To illustrate that, we consider now the case $n=1$. This could give arise
materials with a double 6-polygon structure. This involves two 6-polygons
generated by two length parameters $a_{1}$ and $a_{2}$ required by 
\begin{equation}
a_{2}=\sqrt{3}a_{1}.
\end{equation}%
The associated Lie symmetry is constrained by 
\begin{equation}
\langle \alpha _{1},\alpha _{2}\rangle =-3, \quad \langle \alpha _{2},\alpha
_{1}\rangle =-1, \qquad |\Delta|=12, \qquad |\alpha _{2}|=\sqrt{3}|\alpha
_{1}|.  \label{8atoms}
\end{equation}%
The corresponding material structure with the flat geometry can be derived
from the fact 6-polygons tessellate the full plane producing a new crystal
by mixing the $a(1\times 1)$ and $a(\sqrt{3}\times \sqrt{3})$ configurations
rotated by the angle 
\begin{equation}
\phi_1^6=\frac{\pi}{6}.
\end{equation}
It turns out that we can do something similar for $n>1$. For this case, one
can consider Cartan matrix  
\begin{equation}
A=%
\begin{pmatrix}
2 & -(2n+1) \\ 
-1 & 2%
\end{pmatrix}%
\end{equation}
with the following constraint 
\begin{equation}
4-(2n+1)<0.
\end{equation}
At first sight it is not obvious how to get the corresponding Dynking
diagrams. On the basis of the known results of $G_2$ Lie symmetry, it is
natural to think about a particular geometry involving a central node being
connected with $2n+1$ legs sharing similar properties with the star graph
containing a central node connected with $2n+1$ external ones. The above
Cartan matrix could be obtained by using the standard techniques of folding%
\cite{14,15}. This operation permutes nodes by an outer-automorphism group
leaving the star graph invariant. Precisely, this geometric operation folds
the nodes being permuted by $S_{2n+1}$ symmetric permutation group. Indeed,
this provides a graph involving two nodes connected by $2n+1$ lines. This
extended symmetry may offer a double hexagonal structure based on two
6-polygons with different size being rotated by $\phi_n^6=\frac{n\pi}{6}$.

\section{Conclusion and discussions}

In this work, we have presented a new take on 2D materials based on double
lattice structures involving periodic polygons. In tessellated periodic
representation, two periodic $k$-polygons of unequal side lengths at certain
angles fixed by the underling discrete symmetries are involved. This
procedure has provided two superstructures on the same atomic sheet
generated by two length parameters $a_{1}$ and $a_{2}$ and rotated by the
angle $\frac{n\pi }{k}$. We believe that these geometrical configurations
could be exploited to engineer materials by doubling the number of unit cell
atoms. To support this conjecture, we have linked such structures with Lie
symmetries including finite and indefinite ones. This correspondence between
the Dynkin diagrams and 2D materials could be explored in both directions.
In particular, one may use 2D materials to revise partial results on
indefinite Lie algebras being still an open problem. This may have a new
method to deal with the theory of indefinite Lie symmetries from 2D material
physical point of view.\newline
This work comes up with certain questions. A natural one is to consider
numerical calculations to check the validity of 2D materials with double
periodic structures.

\begin{acknowledgement}
The authors are grateful to their families for support.
\end{acknowledgement}

\end{document}